\documentstyle[12pt]{article}
\begin{document}

\centerline{\Large\bf Modified Palatini action that gives the Einstein-Maxwell theory}
\vskip .7in
\centerline{Dan N. Vollick}
\vskip .2in
\centerline{Irving K. Barber School of Arts and Sciences}
\centerline{University of British Columbia Okanagan}
\centerline{3333 University Way}
\centerline{Kelowna, B.C.}
\centerline{Canada}
\centerline{V1V 1V7}
\vskip 0.5in
\centerline{\bf\large Abstract}
\vskip 0.5in
\noindent

The actions for bosonic fields typically contain terms quadratic in the derivatives of the fields. This is not the case in the Palatini
approach to general relativity. The action does not contain any derivatives of the metric and it only contains terms linear in the derivatives of the connection.
In general relativity the covariant derivative of the metric vanishes, so it is not possible to include such terms in the action. However, in more general theories this is not the case.

In this paper I consider an action which is quadratic in the derivatives of the metric and connection and show that it leads to the coupled
Einstein-Maxwell theory or the coupled Einstein-Proca theory with the antisymmetric part of the Ricci tensor playing the role of the electromagnetic field strength.

\newpage
\section{Introduction}
In the standard Palatini approach to general relativity the action is taken to be
\begin{equation}
S=-\frac{1}{2\kappa}\int g^{\mu\nu}R_{\mu\nu}(\Gamma)\sqrt{g}d^4x
\end{equation}
where $\kappa=8\pi G$ and
\begin{equation}
R_{\mu\nu}(\Gamma)=\partial_{\nu}\Gamma^{\alpha}_{\mu\alpha}-\partial_{\alpha}\Gamma^{\alpha}_{\mu\nu}+\Gamma^{\alpha}_{\beta\mu}
\Gamma^{\beta}_{\alpha\nu}-\Gamma^{\beta}_{\mu\nu}\Gamma^{\alpha}_{\beta\alpha}.
\end{equation}
The metric and connection are varied independently and the resulting field equations are the Einstein equations with the connection given by
the Christoffel symbol. The Palatini action differs from the actions that appears in the standard model. In the standard model the
actions for bosonic fields contain quadratic terms in the derivatives of the fields. In the Palatini action there are no derivatives of the metric and no quadratic terms in the derivatives of the connection, only linear ones. If the connection is the Christoffel symbol, as it is in general relativity, the covariant derivative of the metric vanishes so the action cannot contain terms such terms. However, in more general theories the connection is not necessarily given by the Christoffel symbol and the covariant derivative of the metric does not
vanish. Thus, we can consider adding
\begin{equation}
\sqrt{g}\,(\nabla_{\mu}g_{\alpha\beta})(\nabla^{\mu}g^{\alpha\beta})
\label{metric}
\end{equation}
to the Lagrangian. Here round brackets denote symmeterization and square brackets denote antisymmeterization.
One can also consider adding terms quadratic in the derivatives of the connection such as
\begin{equation}
\sqrt{g}\,\,R^2,\;\;\;\;\; \sqrt{g}\,R_{(\mu\nu)}R^{(\mu\nu)},\;\;\;\;\; and \;\;\;\;\; \sqrt{g}\,R_{[\mu\nu]}R^{[\mu\nu]}
\label{RR}
\end{equation}
to the action. Throughout this paper the connection will be taken to be symmetric.

In this paper I show that the action
\begin{equation}
S=-\int \left[\frac{1}{2\kappa}g^{\mu\nu}R_{\mu\nu}+ R_{[\mu\nu]}R^{[\mu\nu]}+\frac{\beta}{4\kappa}
(\nabla_{\mu}g_{\alpha\beta})(\nabla^{\mu}g^{\alpha\beta})\right]\sqrt{g}d^4x
\end{equation}
gives the coupled Einstein-Maxwell field equations if $\beta=\frac{1}{4}(1\pm\sqrt{13})$ and the coupled Einstein-Proca field equations otherwise. Here I have taken the signature of the metric to be $(-,+,+,+)$.
The antisymmetric part of the Ricci tensor turns out to be proportional to the electromagnetic field tensor.
In addition I show that the only quadratic term in the curvature that can be added to the action, without changing the structure of the field equations, is $R_{[\mu\nu]}R^{[\mu\nu]}$. If other quadratic terms,
such as the first two terms in (\ref{RR}), are added to the action the equation for the connection changes from an algebraic equation to a
differential equation.
I also show how point charges can be included in the theory.

\section{The Field Equations}
As discussed in the introduction I will consider an action of the form
\begin{equation}
S=-\int \left[\frac{1}{2\kappa}g^{\mu\nu}R_{\mu\nu}+\alpha R_{[\mu\nu]}R^{[\mu\nu]}+\frac{\beta}{4\kappa}
(\nabla_{\mu}g_{\alpha\beta})(\nabla^{\mu}g^{\alpha\beta})+L_M\right]\sqrt{g}d^4x
\end{equation}
where $L_M$ is the matter Lagrangian. An action containing the $\sqrt{g}R_{[\mu\nu]}R^{[\mu\nu]}$ term but with $\beta=0$ was discussed by Buchdahl \cite{Bu1}.
Varying the action with respect to the metric gives
\begin{equation}
G_{(\mu\nu)}(\Gamma)=-\kappa \left[T_{\mu\nu}^{(F)}+T_{\mu\nu}^{(g)}+T_{\mu\nu}^{(M)}\right]
\label{EFE}
\end{equation}
where
\begin{equation}
T_{\mu\nu}^{(F)}=\alpha\left[F_{\mu\alpha}F_{\nu}^{\;\;\alpha}-\frac{1}{4}g_{\mu\nu}F^{\alpha\beta}F_{\alpha\beta}\right],
\end{equation}
\begin{equation}
T_{\mu\nu}^{(g)}=\frac{\beta}{2\kappa}\left\{(\nabla_{\mu}g_{\alpha\beta})(\nabla_{\nu}g^{\alpha\beta})-\frac{1}{2}g_{\mu\nu}(\nabla_{\lambda}g_{\alpha\beta})
(\nabla^{\lambda}g^{\alpha\beta})+\frac{1}{\sqrt{g}}\left[g_{\mu\alpha}g_{\nu\beta}\nabla_{\lambda}(\sqrt{g}\;\nabla^{\lambda} g^{\alpha\beta})
-\nabla_{\lambda}(\sqrt{g}\,\nabla^{\lambda} g_{\mu\nu})\right]\right\},
\end{equation}
$G_{\mu\nu}$ is the Einstein tensor, $T_{\mu\nu}^{(M)}$ is the matter energy-momentum tensor, $F_{\mu\nu}=-2R_{[\mu\nu]}=\partial_{\mu}\Gamma_{\nu}-\partial_{\nu}\Gamma_{\mu}$ and $\Gamma_{\mu}=\Gamma^{\alpha}_{\mu\alpha}$.

Varying the action with respect to the connection gives
\begin{equation}
\nabla_{\alpha}(\sqrt{g}g^{\mu\nu})-\Lambda^{(\mu}\delta^{\nu)}_{\alpha}-2\beta g_{\alpha\beta}\nabla^{(\mu}\left(\sqrt{g}g^{\nu)\beta}\right)=0
\label{Gamma}
\end{equation}
where
\begin{equation}
\Lambda^{\mu}=\nabla_{\beta}(\sqrt{g}g^{\beta\mu})+2\alpha\kappa\sqrt{g}\,\tilde{\nabla}_{\beta}F^{\beta\mu}-2\beta\nabla^{\mu}\sqrt{g},
\label{Lambda1}
\end{equation}
$\tilde{\nabla}_{\beta}$ denotes the metric compatible covariant derivative and I have assumed that $L_M$ is independent of the connection.
Contracting over $\mu$ and $\alpha$ gives
\begin{equation}
\Lambda^{\mu}=\frac{2}{5}\left[(1-\beta)\nabla_{\alpha}(\sqrt{g}g^{\alpha\mu})-\beta g_{\alpha\beta}\nabla^{\mu}(\sqrt{g}g^{\alpha\beta})\right].
\label{Lambda2}
\end{equation}
Substituting this into (\ref{Gamma}) gives
\begin{equation}
\nabla_{\alpha}(\sqrt{g}g^{\mu\nu})-2\beta g_{\alpha\beta}\nabla^{(\mu}\left(\sqrt{g}g^{\nu)\beta}\right)-\frac{2}{5}
(1-\beta)\nabla_{\beta}\left(\sqrt{g}g^{\beta(\mu}\right)\delta^{\nu)}_{\alpha}+\frac{2}{5}\beta g_{\lambda\beta}\delta^{(\mu}_{\alpha}
\nabla^{\nu)}\left(\sqrt{g}g^{\lambda\beta}\right)=0.
\label{Gamma2}
\end{equation}
Since the trace of the left hand side of (\ref{Gamma2}) vanishes there are four too few equations and the system is underdetermined. Thus, we
expect four arbitrary functions in the solution. The unique solution to (\ref{Gamma2}) is given by (see appendix)
\begin{equation}
\nabla_{\alpha}(\sqrt{g}g^{\mu\nu})=-\sqrt{g}\left[\delta^{\mu}_{\alpha}V^{\nu}+\delta^{\nu}_{\alpha}V^{\mu}+2\beta g^{\mu\nu}V_{\alpha}
\right]
\label{V}
\end{equation}
and
\begin{equation}
\Gamma^{\alpha}_{\mu\nu}= \left\{\begin{array}{cc}
                           \;\alpha & \\
                            \hspace {-0.1in}\mu&\hspace {-0.3in}\nu
                          \end{array}\hspace {-0.12in}\right\}
-\left(\frac{3}{2}+\beta\right)g_{\mu\nu}V^{\alpha}+\left(\frac{1}{2}+\beta\right)\left(\delta^{\alpha}_{\mu}V_{\nu}+
\delta^{\alpha}_{\nu}V_{\mu}\right)
\label{Gam}
\end{equation}
where $V^{\mu}$ is an arbitrary vector.
Equation (\ref{Gam}) implies that
\begin{equation}
\nabla_{\mu}\sqrt{g}=-(1+4\beta)\sqrt{g}\,V_{\mu}.
\label{detg}
\end{equation}
From (\ref{Lambda1}), (\ref{Lambda2}), (\ref{V}) and (\ref{detg}) we find that
\begin{equation}
\tilde{\nabla}_{\beta}F^{\beta\mu}=-\frac{1}{2\alpha\kappa}\left(4\beta^2-2\beta-3\right)V^{\mu}
\label{Max}
\end{equation}
and
\begin{equation}
F_{\mu\nu}=(1+4\beta)\left(\partial_{\mu}V_{\nu}-\partial_{\nu}V_{\mu}\right).
\end{equation}
Thus if
\begin{equation}
\beta=\frac{1}{4}\left(1\pm\sqrt{13}\right)
\label{beta}
\end{equation}
and we define
\begin{equation}
A_{\mu}=(1+4\beta)V_{\mu}
\end{equation}
equation (\ref{Max}) becomes Maxwell's equations for the vector potential $A_{\mu}$.
To obtain the energy-momentum tensor in Maxwell's theory we take $\alpha=1$.
If $\beta$ is not given by (\ref{beta}) the field equations are Proca's equations with a real mass if
$\frac{(4\beta^2-2\beta-3)}{1+4\beta}<0$ and imaginary mass if $\frac{(4\beta^2-2\beta-3)}{1+4\beta}>0$. If $\beta=-\frac{1}{4}$ equation (18) implies that $F_{\mu\nu}=0$. Equations (17) and (20) then imply that $V^{\mu}$ and $A^{\mu}$ equal zero. In this case the connection is the Christoffel symbol and the field equations are the standard Einstein equations (see equation (24)). Thus, if $\beta=-\frac{1}{4}$ the theory reduces to general relativity.

The energy-momentum tensor $T_{\mu\nu}^{(g)}$ can be now written as
\begin{equation}
T_{\mu\nu}^{(g)}=\frac{\beta}{\kappa}\left[(1+2\beta)g_{\mu\nu}(\tilde{\nabla}_{\alpha}V^{\alpha})-(\tilde{\nabla}_{\mu}V_{\nu}+
\tilde{\nabla}_{\nu}V_{\mu})
+(7+4\beta-8\beta^2)V_{\mu}V_{\nu}-\frac{1}{2}(1-8\beta^2)g_{\mu\nu}(V_{\alpha}V^{\alpha})\right].
\end{equation}

\noindent
It will be useful to write the field equations in terms of $\tilde{G}_{\mu\nu}$, which is defined in terms of the Christoffel symbol. The relationship between the Ricci tensors is
\begin{equation}
R_{\mu\nu}(\Gamma)=\tilde{R}_{\mu\nu}-\tilde{\nabla}_{\alpha}H^{\alpha}_{\mu\nu}+\tilde{\nabla}_{\nu}H^{\alpha}_{\alpha\mu}
-H^{\alpha}_{\alpha\beta}H^{\beta}_{\mu\nu}+H^{\alpha}_{\mu\beta}H^{\beta}_{\alpha\nu}
\end{equation}
where
\begin{equation}
H^{\alpha}_{\mu\nu}=\Gamma^{\alpha}_{\mu\nu}-\left\{\begin{array}{cc}
                           \;\alpha & \\
                            \hspace {-0.1in}\mu&\hspace {-0.3in}\nu
                          \end{array}\hspace {-0.12in}\right\}.
\end{equation}
The Einstein Equations can now be written as
\begin{equation}
\tilde{G}_{\mu\nu}(\Gamma)=-\kappa \left\{T_{\mu\nu}^{(F)}-\frac{(4\beta^2-2\beta-3)}{2\kappa(1+4\beta)}
\left[A_{\mu}A_{\nu}-\frac{1}{2}g_{\mu\nu}(A_{\lambda}A^{\lambda}-2\tilde{\nabla}_{\lambda}A^{\lambda})\right]\right\}-\kappa T_{\mu\nu}^{(M)}.
\label{Einstein}
\end{equation}
For the Proca field equations it follows from (\ref{Max}) that $\tilde{\nabla}_{\lambda}A^{\lambda}=0$ and for Maxwell's theory the the coefficient of the $\tilde{\nabla}_{\lambda}A^{\lambda}$ term
vanishes. The $\tilde{\nabla}_{\lambda}A^{\lambda}$ term can therefore be dropped in (\ref{Einstein}). The energy-momentum tensor in the brackets on the right hand side of (\ref{Einstein}) is the Maxwell energy-momentum tensor if $4\beta^2-2\beta-3=0$ and Proca's energy-momentum tensor otherwise. Equations (\ref{Max}) and (\ref{Einstein}) therefore give either the coupled Maxwell-Einstein equations or the coupled Einstein-Proca equations, depending on the value of $\beta$. It is interesting to note that if (15) is substituted into the action (5) one gets, after dropping a term involving a total derivative, the the Einstein-Hilbert action plus either Proca's or Maxwell's action, depending on the value of $\beta$.

\noindent So far I have only considered adding the quadratic term $R_{[\mu\nu]}R^{[\mu\nu]}$ involving the curvature to the action. This term only contains terms quadratic in the derivatives of $\Gamma_{\mu}$ not in the derivatives of the full connection. Now consider adding a term involving $R^2$ to the action. In fact it will be possible to analyze the theory for $f(R)$ gravity \cite{Fa1} where the gravitational Lagrangian
$-\frac{1}{2\kappa}\sqrt{g}\,R$ is replaced by $-\frac{1}{2\kappa}\sqrt{g}f(R)$. For simplicity I will consider only the matter free equations.
Varying the action with respect to the metric gives
\begin{equation}
f'(R)R_{\mu\nu}-\frac{1}{2}f(R)g_{\mu\nu}=-\kappa\left[T^{(F)}_{\mu\nu}+T^{(g)}_{\mu\nu}\right]
\label{1}
\end{equation}
and varying the action with respect to the connection gives
\begin{equation}
\nabla_{\alpha}\left(\sqrt{g}f'(R)g^{\mu\nu}\right)-\Lambda^{(\mu}\delta^{\nu)}_{\alpha}-2\beta g_{\alpha\beta}\nabla^{(\mu}\left(\sqrt{g}g^{\nu)\beta}\right)=0
\label{2}
\end{equation}
where
\begin{equation}
\Lambda^{\mu}=\nabla_{\beta}\left(\sqrt{g}f'(R)g^{\beta\mu}\right)+2\alpha\kappa\sqrt{g}\,\tilde{\nabla}_{\beta}F^{\beta\mu}-2\beta\nabla^{\mu}\sqrt{g}.
\end{equation}
When $f(R)=R$ these equations give algebraic, not differential, equations for the connection. However, when $f(R)\neq R$ we will see that these equations are second order differential equations for the connection and are therefore very different from the equations in the usual Palatini
approach (a similar issue can arise in generalized Palatini theories that do not contain derivatives of the metric \cite{Vi1})
Since $R$ contains first order derivatives of the connection the field equations (\ref{2}), as they stand, are second order in the connection. In $f(R)$ theories there is a similar problem and it can be resolved by taking the trace of the field equations that correspond to (\ref{1}), solving for $R$ in terms of $T=g_{\mu\nu}T^{\mu\nu}$ and substituting
it into the equations that correspond to (\ref{2}). Here the trace of (\ref{1}) is given by
\begin{equation}
Rf'(R)-2f(R)=-\kappa T^{(g)}
\label{4}
\end{equation}
where $T^{(g)}=g^{\mu\nu}T^{(g)}_{\mu\nu}$. In $f(R)$ theories $T^{(g)}=0$ and the above reduces to an algebraic equation for $R$. Assuming a
solution exists we find that $R$ is a constant and the equation for the connection is algebraic. Here the solution to (\ref{4}) will be of the form
\begin{equation}
R=R\left[T^{(g)}\right].
\end{equation}
For example, if $f(R)=R+\lambda R^2$ the solution is $R=\kappa T^{(g)}$.
Now
\begin{equation}
T^{(g)}=\frac{\beta}{\kappa}\left[\frac{2}{\sqrt{g}}\nabla_{\mu}(\sqrt{g}H^{\mu})+g^{\mu\nu}\left(H^{\alpha}_{\mu\beta}H^{\beta}_{\nu\alpha}
+H^{\lambda}_{\mu\alpha}H^{\beta}_{\nu\rho}g_{\lambda\beta}g^{\alpha\rho}\right)\right]
\end{equation}
showing that $T^{(g)}$ involves first order derivatives of the connection.
Equation (\ref{2}) therefore remains a second order differential equation for the connection and the theory is very different from the usual Palatini theory. A similar problem will occur if we add terms such as $\sqrt{g}\,R_{(\mu\nu)}R^{(\mu\nu)}$ or $\sqrt{g}R_{\alpha\beta\mu\nu}
R^{\alpha\beta\mu\nu}$ to the Lagrangian. Here equation (\ref{Gamma}) is a first order differential equation involving $\tilde{\nabla}_{\beta}F^{\beta\mu}$. This equation can be replaced by the algebraic equation (\ref{Gamma2}) which is underdetermined and whose solution contains an arbitrary vector field plus a differential equation (\ref{Max}) involving this vector field. The differential equation is
either Maxwell's or Proca's equation and can be solved together with Einstein's equations given initial conditions. This approach does not work for $f(R)$ theories.
Thus, given that we add the term $\sqrt{g}\,\left(\nabla_{\mu}g_{\alpha\beta}\right)
\left(\nabla^{\mu}g^{\alpha\beta}\right)$ to the Lagrangian the only quadratic term in the curvature that can be added is $\sqrt{g}\,
R_{[\mu\nu]}R^{[\mu\nu]}$.
Other terms involving derivatives of the metric, such as $\sqrt{g}\,\left(\nabla_{\mu}g_{\alpha\beta}\right)
\left(\nabla^{\alpha}g^{\mu\beta}\right)$ and $\sqrt{g}\,(\nabla_{\mu}g)(\nabla^{\mu}g)/g^2$, could also be added to the action, but I will not consider them here. I have considered the term $\sqrt{g}\,(\nabla_{\mu}g)(\nabla^{\mu}g)/g^2$ on its own in an unpublished article
\cite{Vo2} and showed that one obtains results similar to those obtained here. However, the Lagrangian $\sqrt{g}\,(\nabla_{\mu}g_{\alpha\beta})(\nabla^{\mu}g^{\alpha\beta})$ is certainly more natural than $\sqrt{g}\,(\nabla_{\mu}g)(\nabla^{\mu}g)/g^2$.

Before leaving this section I would like to briefly comment on the relationship of the theory presented in this paper to hybrid theories \cite{Ca1}
and to Weyl's theory \cite{We1}.
In hybrid theories the Lagrangian contains terms involving the Ricci tensor expressed in terms of the metric and terms involving the Ricci tensor expressed in terms of connection. A simple example is given by the Lagragian $L=-\frac{1}{2\kappa}\sqrt{g}\{R(g)+f[R(\Gamma)]\}$. One key difference between the two theories is that the Lagrangians of hybrid theories contain second order derivatives of the metric while the Lagrangian of the theory presented in this paper contains only first order derivatives of the metric.

The connection $\Gamma^{\alpha}_{\mu\nu}$ given in (\ref{Gam}) is similar to the connection that appears in Weyl's theory. However, Weyl's theory differs significantly from the theory presented here. In Weyl's theory the metric is not a physically meaningful quantity. This is due to the fact that the theory admits conformal transformations. Thus, only equivalence classes $\{\lambda g_{\mu\nu}|\lambda$ an arbitrary function$\}$ are physically meaningful in Weyl's theory \cite{Go1}. In contrast the metric is a physically meaningful quantity in this paper.

\section{Sources of the Electromagnetic Field}
In this section I take $\beta=\frac{1}{4}(1\pm\sqrt{13})$ and consider coupling point charges to the electromagnetic field. The
interaction Lagrangian will be taken to be \cite{Vo1}
\begin{equation}
L_I=\sqrt{g}\,\Gamma_{\mu}J^{\mu}
\end{equation}
where
\begin{equation}
J^{\mu}(x^{\alpha})=\frac{1}{\sqrt{g}}\Sigma_{n} e_n\int\frac{dx^{\mu}_n}{d\tau_n}\,\delta^{(4)}(x^{\alpha}-x^{\alpha}_n)d\tau_n
\end{equation}
is the current density for the charges. Here $e_n$ and $\tau_n$ are the charge and proper time along the worldline of the n$^{th}$ particle.

Since $\Gamma_{\mu}$ is not a vector we need to verify that the action is well defined under a coordinate transformation. The field $\Gamma_{\mu}$ transforms as
\begin{equation}
\bar{\Gamma}_{\mu}=\frac{\partial x^{\nu}}{\partial\bar{x}^{\mu}}\Gamma_{\nu}+\frac{\partial}{\partial \bar{x}^{\mu}}\ln\left|
\frac{\partial x}{\partial\bar{x}}\right|
\end{equation}
under a coordinate transformation, where $|\partial x/\partial\bar{x}|$ is the Jacobian of the transformation. This is analogous to a gauge transformation and it is easy to see that
$L_I$ only changes by a total derivative under a coordinate transformation if $\tilde{\nabla}_{\mu}J^{\mu}=0$. This is required by charge conservation indicating that $L_I$ is well behaved under a coordinate transformation.

This interection Lagrangian does not change (\ref{EFE}), since it is independent of $g_{\mu\nu}$. It will, however, change the equations that follow from the variation of the connection. This change amounts to replacing $\tilde{\nabla}_{\beta}F^{\beta\mu}$ in (\ref{Lambda1}) with
$\tilde{\nabla}_{\beta}F^{\beta\mu}+J^{\mu}$. The field equations for $F^{\mu\nu}$ become
\begin{equation}
\tilde{\nabla}_{\beta}F^{\beta\mu}=-J^{\mu},
\label{MFE}
\end{equation}
which are Maxwell's equations with a source term.
The matter action is given by
\begin{equation}
S_M=-\Sigma_nm_n\int\sqrt{-g_{\mu\nu}U^{\mu}_nU^{\nu}_n}d\tau_n
\end{equation}
where $U^{\mu}=\frac{dx_n^{\mu}}{d\tau_n}$.The equations of motion that follow from the interaction and matter actions are
\begin{equation}
\frac{dU_n^{\mu}}{d\tau_n}+\left\{\begin{array}{cc}
                           \;\mu & \\
                            \hspace {-0.1in}\alpha &\hspace {-0.3in}\beta
                          \end{array}\hspace {-0.12in}\right\}
U^{\alpha}_nU^{\beta}_n=e_nF^{\mu}_{\;\;\;\nu}U^{\nu}_n.
\label{point}
\end{equation}
Thus, by the inclusion of the interaction and matter actions the theory describes point charges interacting with the electromagnetic field.

It is interesting to note that it is the Christoffel symbol not $\Gamma^{\alpha}_{\mu\nu}$ that appears on the right hand side of (\ref{point}).
Thus, neutral particles respond only to the Christoffel symbol not to $\Gamma^{\alpha}_{\mu\nu}$. The field equations (\ref{MFE}) for the electromagnetic field also involve only the Christoffel symbol. In Maxwell's theory $A_{\mu}$ is defined only up to a gauge transformation which means that $V_{\mu}$ is also only defined up to a gauge transformation. This tells us that $\Gamma^{\alpha}_{\mu\nu}$ is not uniquely defined. This does not present a problem since the equations of motion for the particles and fields are independent of this ambiguity.

\section{Conclusion}
In this paper I considered adding quadratic terms in the derivatives of the metric and connection to the standard Palatini action. Terms that are quadratic in the derivatives of the fields usually appear in the actions for bosonic fields, so it is natural to include such terms in the Palatini action. The term quadratic in the derivatives of the metric was taken to be $-\frac{\beta}{4\kappa}\sqrt{g}\,(\nabla_{\mu}g_{\alpha\beta})(\nabla^{\mu}g^{\alpha\beta})$ and the term quadratic in the derivatives of the connection was taken to be proportional to $\sqrt{g}\,R_{[\mu\nu]}R^{[\mu\nu]}$. The connection in this theory is given by
\begin{equation}
\Gamma^{\alpha}_{\mu\nu}= \left\{\begin{array}{cc}
                           \;\alpha & \\
                            \hspace {-0.1in}\mu&\hspace {-0.3in}\nu
                          \end{array}\hspace {-0.12in}\right\}
-\left(\frac{3}{2}+\beta\right)g_{\mu\nu}V^{\alpha}+\left(\frac{1}{2}+\beta\right)\left(\delta^{\alpha}_{\mu}V_{\nu}+
\delta^{\alpha}_{\nu}V_{\mu}\right)
\end{equation}
where $V_{\mu}$ is an arbitrary vector field.
If $\beta=\frac{1}{4}(
1\pm\sqrt{13})$ the field equations describe the free electromagnetic field coupled to gravity with the vector field proportional to the vector potential and with $R_{[\mu\nu]}$ proportional to the electromagnetic field strength tensor. For other values of $\beta$ the field equations are the Einstein-Proca equations.

I showed that if the term $-\frac{1}{4\beta\kappa}\sqrt{g}\,(\nabla_{\mu}g_{\alpha\beta})(\nabla^{\mu}g^{\alpha\beta})$ is added to the Lagrangian it is not possible to add additional terms quadratic in the curvature without changing the equation for the connection from an algebraic equation to a differential equation. I also showed that sources to the electromagnetic field can be included by adding
$\sqrt{g}\,\Gamma_{\mu}J^{\mu}$ to the Lagrangian.
\section{Appendix}
In this appendix equation (\ref{Gamma2}) is solved. Substituting
\begin{equation}
\Gamma^{\alpha}_{\mu\nu}=\left\{\begin{array}{cc}
                           \;\alpha & \\
                            \hspace {-0.1in}\mu&\hspace {-0.3in}\nu
                          \end{array}\hspace {-0.12in}\right\}
                          +H^{\alpha}_{\mu\nu}
\end{equation}
into (\ref{Gamma2}) gives
\begin{equation}
(1-\beta)(H^{\mu\alpha\nu}+H^{\nu\alpha\mu})-H^{\alpha}g^{\mu\nu}+\frac{3}{5}\beta(H^{\mu}g^{\alpha\nu}+H^{\nu}g^{\alpha\mu})-2\beta H^{\alpha\mu\nu}-\frac{1}{5}(1-\beta)g_{\beta\rho}(H^{\mu\beta\rho}g^{\alpha\nu}+H^{\nu\beta\rho}g^{\alpha\mu})=0
\end{equation}
where $H^{\mu\alpha\beta}=g^{\alpha\lambda}g^{\beta\sigma}H^{\mu}_{\lambda\sigma}$. Now consider this equation and two cyclic permutations. Adding the first two and subtracting the third gives
\begin{equation}
H^{\mu}_{\alpha\beta}-\beta g^{\mu\lambda}(g_{\alpha\sigma}H^{\sigma}_{\lambda\beta}+g_{\beta\sigma}H^{\sigma}_{\alpha\lambda})+(1+\frac{6}{5}\beta)H^{\mu}g_{\alpha\beta}-
H_{\alpha}\delta^{\mu}_{\beta}-H_{\beta}\delta_{\alpha}^{\mu}-\frac{2}{5}(1-\beta)(H^{\mu}_{\beta\sigma}g^{\beta\sigma})g_{\alpha\beta}=0
\label{H}
\end{equation}
 where I have lowered two of the indices. Contracting over $\mu$ and $\alpha$ gives
 \begin{equation}
 \left(1+\frac{2}{5}\beta\right)H^{\mu}+\frac{1}{5}\left(1+4\beta\right)\left(H^{\mu}_{\alpha\beta}g^{\alpha\beta}\right)=0.
 \label{trace}
 \end{equation}
 Now using (\ref{trace}) equation (\ref{H}) can be written as
 \begin{equation}
 H^{\mu}_{\;\;\,\alpha\beta}-\beta\left(H_{\alpha\;\;\beta}^{\;\;\mu}+H_{\beta\alpha}^{\;\;\;\;\,\mu}\right)=\Omega^{\mu}_{\;\;\,\alpha\beta}
 \label{H2}
 \end{equation}
 where
 \begin{equation}
 \Omega^{\mu}_{\;\;\,\alpha\beta}=\frac{1}{2}\left[H_{\alpha}\delta^{\mu}_{\beta}+H_{\beta}\delta^{\mu}_{\alpha}-\left(
 \frac{3+4\beta+4\beta^2}{1+4\beta}\right)H^{\mu}g_{\alpha\beta}\right].
 \label{Omega}
 \end{equation}
 Lowering $\mu$, raising $\alpha$ and then interchanging $\alpha$ and $\mu$ in (\ref{H2}) gives
 \begin{equation}
 H_{\alpha\;\;\beta}^{\;\;\mu}-\beta\left(H^{\mu}_{\;\;\,\alpha\beta}+H_{\beta\;\;\alpha}^{\;\;\mu}\right)=\Omega_{\alpha\;\;\beta}^{\;\;\mu}.
 \label{H3}
 \end{equation}
 Lowering $\mu$, raising $\beta$ and then interchanging $\beta$ and $\mu$ in (\ref{H2}) gives
 \begin{equation}
 H_{\beta\alpha}^{\;\;\;\;\,\mu}-\beta\left(H_{\alpha\;\;\beta}^{\;\;\mu}+H^{\mu}_{\;\;\,\alpha\beta}\right)=\Omega_{\beta\alpha}^{\;\;\;\;\,\mu}.
 \label{H4}
 \end{equation}
Subtracting equation (\ref{H3}) from equation (\ref{H2}) gives
 \begin{equation}
 (1+\beta)\left(H^{\mu}_{\;\;\,\alpha\beta}-H_{\alpha\;\;\beta}^{\;\;\mu}\right)=\Omega^{\mu}_{\;\;\,\alpha\beta}-\Omega_{\alpha\;\;\beta}^{\;\;\mu}
 \label{H5}
 \end{equation}
 and adding $\beta$ times equation (\ref{H4}) to equation (\ref{H2}) gives
 \begin{equation}
 (1-\beta^2)H^{\mu}_{\;\;\,\alpha\beta}-\beta(1+\beta)H_{\alpha\;\;\beta}^{\;\;\mu}=
 \Omega^{\mu}_{\;\;\,\alpha\beta}+\beta\Omega_{\beta\alpha}^{\;\;\;\;\,\mu}.
 \label{H6}
 \end{equation}
 Subtracting $(1-\beta)$ times equation (\ref{H5}) from equation (\ref{H6}) gives
 \begin{equation}
 (1+\beta)(1-2\beta)H_{\alpha\;\;\beta}^{\;\;\mu}=\beta\left(\Omega^{\mu}_{\;\;\,\alpha\beta}+\Omega_{\beta\alpha}^{\;\;\;\;\,\mu}\right)
 +(1-\beta)\Omega_{\alpha\;\;\beta}^{\;\;\mu}.
 \end{equation}
 Using (\ref{Omega}) one finds that
 \begin{equation}
 H^{\mu}_{\alpha\beta}=\frac{(1+2\beta)}{2(1+4\beta)}\left[H_{\alpha}\delta^{\mu}_{\beta}+H_{\beta}\delta^{\mu}_{\alpha}\right]-
 \frac{(\frac{3}{2}+\beta)}{(1+4\beta)}H^{\mu}g_{\alpha\beta}.
 \end{equation}
 This is a solution to (\ref{Gamma2}) for any $H^{\mu}$. If we set
 $H^{\mu}=(1+4\beta)V^{\mu}$ we obtain (\ref{Gam}) and the proof is complete.
\section*{Acknowledgements}
This research was supported by the  Natural Sciences and Engineering Research
Council of Canada.

\end{document}